\documentclass[12pt]{iopart}

\usepackage{graphicx,color}           
\usepackage{gensymb}                  
\usepackage{textcomp}                 
\usepackage{acronym}                  
\usepackage{subfig}

\begin{document}

\title[Huge TAMR in (Ga,Mn)As nanoconstrictions]{Huge tunnelling anisotropic magnetoresistance in (Ga,Mn)As nanoconstrictions}

\author{A D Giddings$^1$, O N Makarovsky$^1$, M N Khalid$^2$, S Yasin$^3$, K W Edmonds$^1$, R P Campion$^1$, J Wunderlich$^{2,4}$, T Jungwirth$^{4,1}$, D A Williams$^2$, B L Gallagher$^1$ and  C T Foxon$^1$}

\address{$^1$ School of Physics and Astronomy, University of Nottingham, Nottingham NG7 2RD, UK}
\address{$^2$ Hitachi Cambridge Laboratory, Cambridge CB3 0HE, UK}
\address{$^3$ Microelectronics Research Centre, Cavendish Laboratory, University of Cambridge, CB3 0HE, UK}
\address{$^4$ Institute of Physics ASCR, Cukrovarnick\'a 10, 162 53 Praha 6, Czech Republic}

\begin{abstract}
We report large anisotropic magnetoresistance (AMR) behaviours in single lateral (Ga,Mn)As nanoconstriction of up to 1300\%, along with large multistable telegraphic switching. The nanoconstriction devices are fabricated using high-resolution electron beam lithography of a 5 nm thick (Ga,Mn)As epilayer. The unusual behaviour exhibited by these devices is discussed in the context of existing theories for enhanced AMR ferromagnetic semiconductor nanoscale devices, particularly with regard to the dependence on the magnetotransport of the bulk material. We conclude that our results are most consistent with the Coulomb blockade AMR mechanism.
\end{abstract}

\ead{ppxadg@nottingham.ac.uk}
\submitto{\NJP}

\maketitle

\section{Introduction}

In conductors there can exist many forms of \ac{MR}, whereby the electrical resistance is modified via the application of a magnetic field. When the size of a \ac{MR} is a function of the angle between the magnetisation and direction of flow of carriers or crystallographic axes it is known as either the non-crystalline or crystalline \ac{AMR}, respectively \cite{mcguire_anisotropic_1975,rushforth_anisotropic_2007}. The origin of this effect is the \ac{SOC}.
We shall refer to this well known \ac{AMR} in the Ohmic regime as the \ac{NAMR}.

When a non-Ohmic tunnelling regime is considered instead, a much more dramatic effect known as \ac{TAMR} can occur. This is caused by the dependence of the tunnelling density of states on the direction of the magnetisation of the material;
thus the tunnelling probabilities can be directly manipulated with the application of a magnetic field, resulting in large magnetoresistance effects.
This was demonstrated initially in vertical structures based on the \ac{DMS} (Ga,Mn)As \cite{gould_tunneling_2004,ruster_very_2005}. Shortly after, we reported data consistent with \ac{TAMR} in lateral nanoconstriction devices \cite{giddings_large_2005}.
As a consequence of this, it was predicted that \ac{TAMR} could be a generic property of tunnel devices with ferromagnetic contacts \cite{shick_prospect_2006}. Since then, the \ac{TAMR} phenomenon has also been reported in transition metal tunnel junction systems \cite{bolotin_anisotropic_2006,gao_bias_2007,shi_temperature_2007,moser_tunneling_2007,park_tunneling_2008}.

In further work on (Ga,Mn)As lateral tunnelling devices another novel magnetoresistance effect was reported,
the so-called \ac{CBAMR} \cite{wunderlich_coulomb_2006}.
The origin of \ac{CBAMR} is anisotropic shifts in the Fermi energy with respect to magnetisation in an inhomogeneous system. This is achieved by patterning a nanoscale \ac{SET} type structure from ferromagnetic material with strong \ac{SOC}.
This observation occurred in an accidentally inhomogeneous constriction with an associated gate that allowed tuning of the local electrostatic conditions; as a result of the patterning the necessary inhomogeneity was created in the form of extremely low capacitance ``islands'' isolated from the rest of the structure by a tunnel barrier.
In systems with strong \ac{SOC}, such as (Ga,Mn)As, magnetisation rotation can cause large changes in the electronic configuration. As a result of the non-uniform local carrier concentration in these structures, changes in the magnetisation orientation causes differential changes in the chemical potential of the nanoscale island and leads.
The Gibbs free energy associated with transmission of charge through the island can be written as a function of these different chemical potentials and as such is dependent on the magnetisation. Furthermore, the difference in the chemical potential between the island and the leads is of a similar order to the single-electron charging energy \cite{wunderlich_coulomb_2006}, resulting in potentially dramatic changes in conductivity. 

A possible third mechanism for large magnetoresistance effects in \ac{DMS} tunnel devices has since also been suggested, in the form of a magnetisation orientation induced metal-insulator transition \cite{pappert_magnetization-switched_2006}. This can occur when a high localisation of carriers, such as at low temperatures or in highly depleted regions, causes transport to go from a diffusive to an Efros-Shklovskii hopping regime. If the structure is therefore close to a metal-insulator transition, and is highly anisotropic due to the strong \ac{SOC}, then changing the magnetisation orientation could trigger the transition.

Putting this in the context of the previous two magnetoresistance effects, it is interesting to note that the ultra-thin (Ga,Mn)As films similar to those used in~\cite{giddings_large_2005,wunderlich_coulomb_2006} become very resistive and exhibit 
hopping-like conductivity at very low temperatures ($T < 4$ K). Additionally, they have unusually strong magnetocrystalline anisotropies \cite{rushforth_amr_2006}, making this effect of interest in regard to these kinds of lateral structures.

Bearing in mind the diverse mechanisms for magnetoresistances that non-Ohmic devices in ferromagnets with strong \ac{SOC} can acquire, in this paper we demonstrate further evidence of extremely large effects in laterally defined (Ga,Mn)As nanoconstriction devices and so cast further light on these issues.

\section{Experimental method}

The device is fabricated from a 5 nm thick (Ga$_{0.94}$,Mn$_{0.06}$)As epilayer grown on a (100) GaAs substrate by low-temperature ($230^\circ$C) molecular beam epitaxy. Beneath the (Ga,Mn)As layer there is a 25 nm AlAs layer. The as-grown sheet resistivity was $90 ~\micro\Omega\,$m at room temperature, although at low temperatures ($T \sim 4$ K) the sample would become insulating. After annealing at $170^\circ$C for 8 hours the material's room temperature resistivity was $51 ~\micro\Omega\,$m and $170 ~\micro\Omega\,$m at 4.2 K, with a Curie temperature of 120 K. This change is the result of the removal of interstitial Manganese \cite{edmonds_mn_2004}.

Sample fabrication was carried out on as-grown material via high resolution electron beam lithography using a \ac{PMMA} positive resist. This was developed using a DI water:IPA solution and etching was achieved with a silicon tetrachloride reactive ion dry etch. During the fabrication process the sample was exposed to annealing level temperatures. Post-processing resistance was about 4 times greater than that of the annealed material, although there was some inhomogeneity between devices.

\begin{figure}
  \centering
  \includegraphics[width=0.7\textwidth]{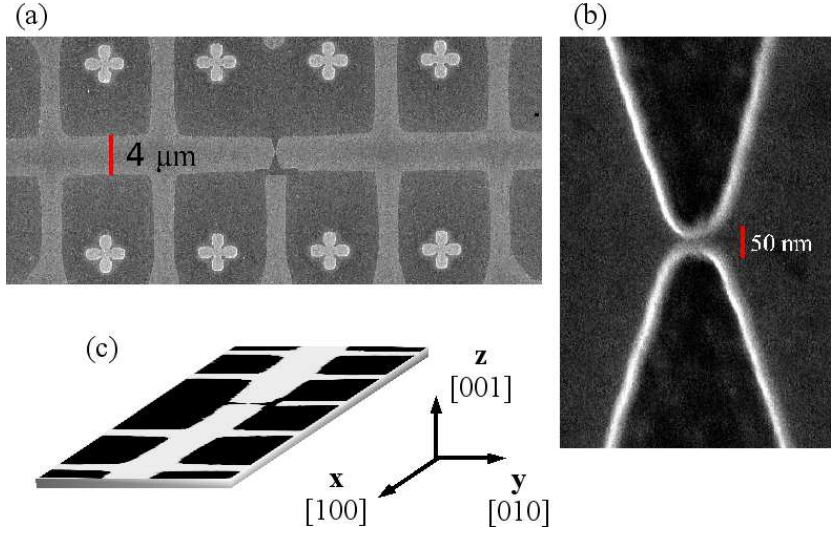}
  \caption{Scanning electron microscope micrographs of the device, showing (a) the hall bar geometry of the device and the positioning of the constriction within it, and (b) a close up image of the nanoconstriction showing it to have a width of about 30 nm. The orientation of the hall bar with respect to the crystalline axes is shown in cartoon (c).}
  \label{SEM}
\end{figure}

The device consists of a Hall bar type structure with a single nanoconstriction, as shown in figure~\ref{SEM}(a). The bar is aligned along the [100] cubic axis, which we shall denote as the ${\bf x}$ axis. The perpendicular in-plane axis is ${\bf y}$ and the perpendicular out-of-plane axis is ${\bf z}$ (see figure~\ref{SEM}(c)). Scanning electron microscope measurements of the nanoconstriction, shown in figure~\ref{SEM}(b), estimate it to have a physical width of about 30 nm. However, carrier depletion and interface effects will make the effective width of the channel smaller. Non-linear current-voltage ($I$-$V$) characteristics develop below $\sim 4$ K are consistent with the development of tunnel barriers or hopping conduction.

Sample measurement was carried out in a He cryostat down to 1.5 K.
An external magnet capable of fields up to 0.7 T could be rotated 180$^\circ$ in the ${\bf x}$-${\bf y}$ plane around the sample. Additionally, the ${\bf z}$ axis of the sample could be rotated by up to 180$^\circ$ with respect to the field. This provides the possibility for any 3D angle of the applied magnetic field to the sample. Using a 4-point sensing measurement scheme, the potential across the constriction is kept constant.

\section{Results and discussion}

\begin{figure}
  \centering%
  \subfloat[]%
    {\includegraphics[width=0.45\textwidth]{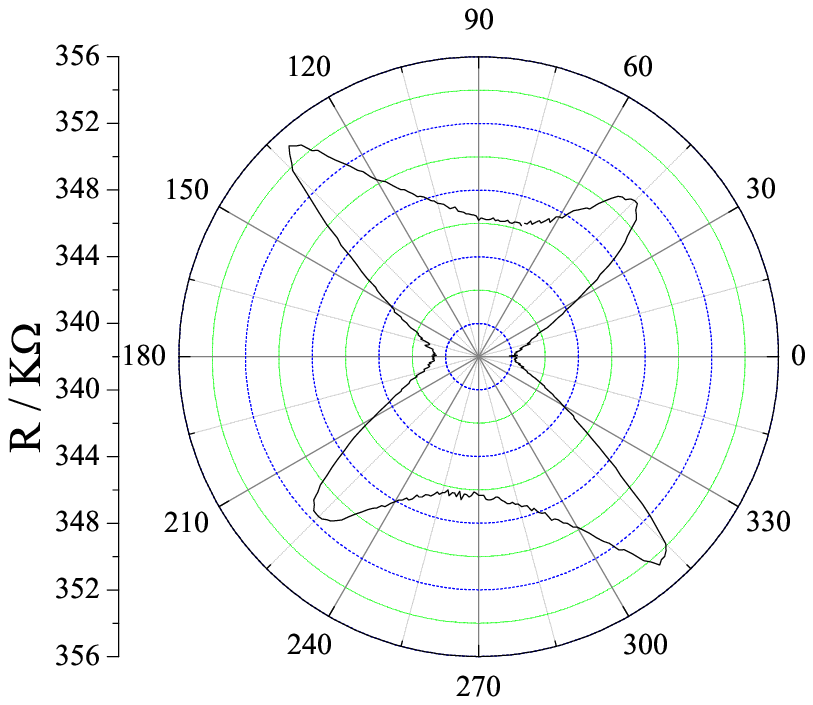}}%
  \subfloat[]%
    {\includegraphics[width=0.45\textwidth]{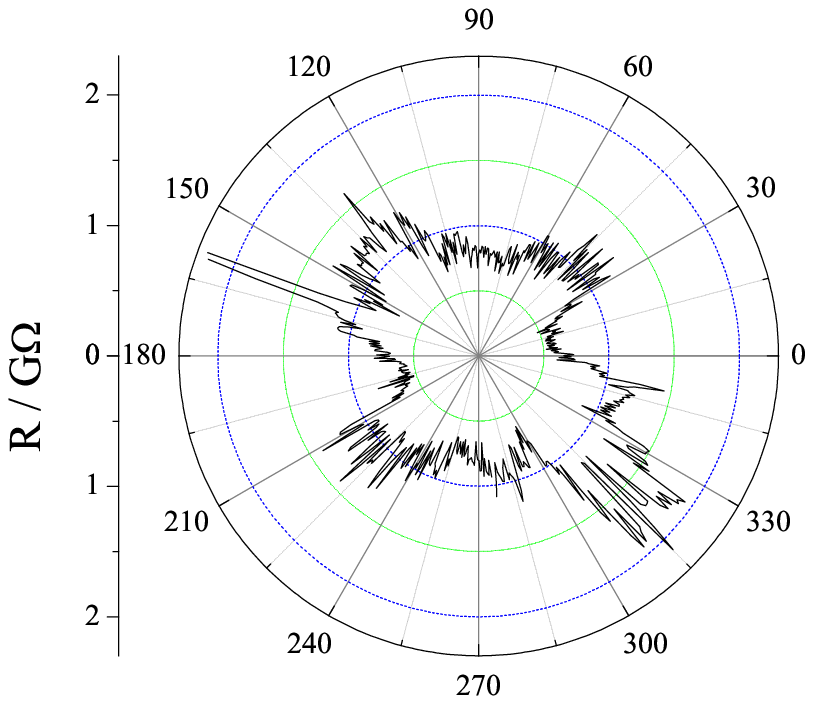}}%
  \caption{Measurements as the sample is rotated in the ${\bf x}$-${\bf y}$ plane in a 0.2 T field. The angle given is between the current and field. (a) Bulk material at 4.2 K. (b) The nanoconstriction at 1.5 K.}%
  \label{rotations}
\end{figure}

Figure~\ref{rotations} compares the measured resistance of the unpatterned section of the Hall bar, at a constant 1 \micro A current, with that of the constriction, with an excitation of 40 mV, as a magnetic field of 0.2 T is rotated in the ${\bf x}$-${\bf y}$ plane of the epilayer. This field strength is greater than that at which hysteresis is observed: putting the measurement outside the range of the large hysteretic effects can give a clearer indication of the \ac{AMR}. The constriction shows a much larger \ac{MR} than the unstructured bar, up to $\sim 300$\%.
There is also much greater richness in features in the measured \ac{MR} of the constriction.
At such small sizes these will be strongly influenced by
local fluctuations of electrostatic potential, which could change between thermal cycles or even hysteric field sweeps. Despite this, there are qualitative similarities between the two traces presented in figure~\ref{rotations}(a) and (b); in both parts of the sample the highest and lowest conductances occur along the same orientations, indicating a close link between the anisotropic magnetic properties of the constriction and bulk material.
Since the magnetocrystalline components of the \ac{NAMR} are dominant in these materials \cite{giddings_large_2005} this indicates that the observed \ac{AMR} in the constriction arises from an anisotropic response of the (Ga,Mn)As material as is the case in \ac{TAMR}, \ac{CBAMR} and induced metal-insulator transitions.

\begin{figure}
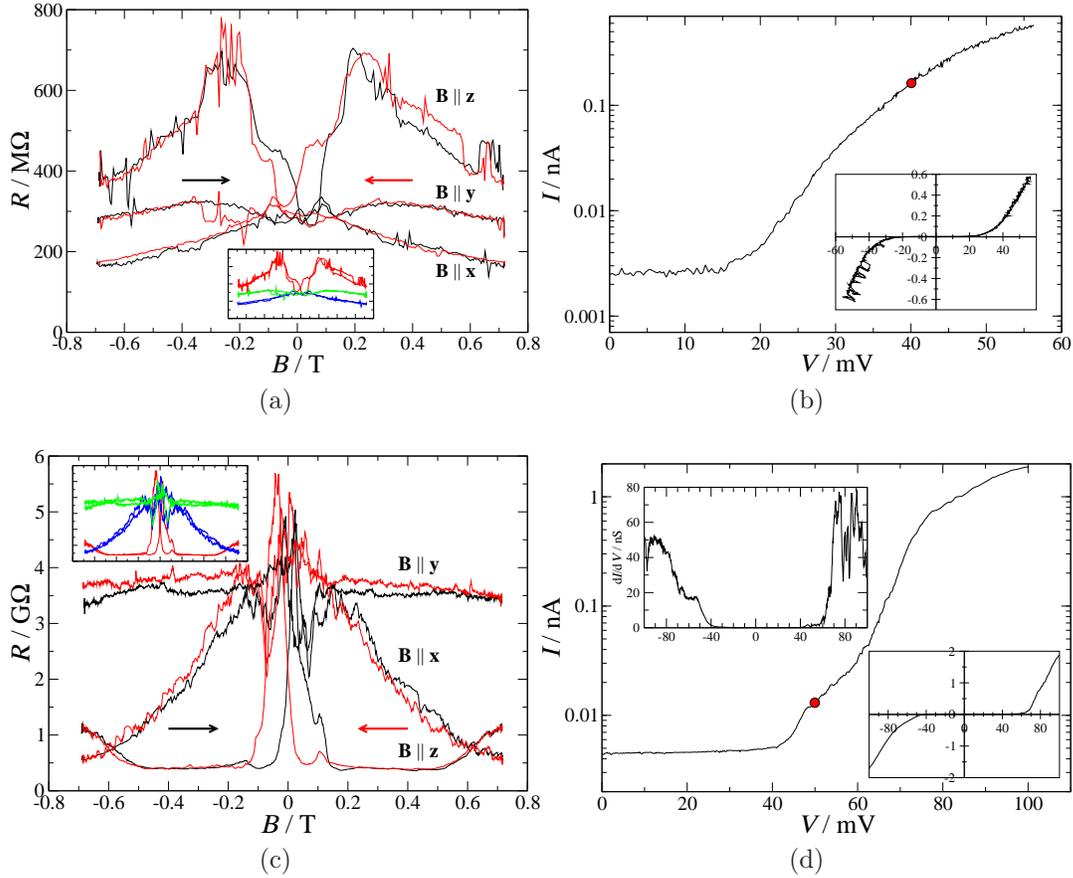

  \centering%
  \subfloat[]%
    {\includegraphics[width=0.45\textwidth]{FIGURES/fig_magtrans_low_res_split}}%
  \subfloat[]%
    {\includegraphics[width=0.45\textwidth]{FIGURES/fig_magtrans_low_res_IV}}\\
  \subfloat[]%
   {\includegraphics[width=0.45\textwidth]{FIGURES/fig_magtrans_high_res_split}}%
  \subfloat[]%
    {\includegraphics[width=0.45\textwidth]{FIGURES/fig_magtrans_high_res_IV2}}%
  \caption{Magnetotransport measurements with the field applied along the three cardinal directions ${\bf x}$, ${\bf y}$ and ${\bf z}$ for a single sample, but during different thermal cycles, and the corresponding $I$-$V$ characteristic at $B = 0$ T. In the magnetotransport figures the black curve is for increasing field and the red curve is for decreasing field. In the inserts the three directions are distinguished by the different colouring. In the $I$-$V$ figures the red dot marks the excitation across the nanoconstriction used for that measurement. The $I$-$V$ curves have been averaged between up and down sweeps except in the case of the insert in (b) so as to preserve the switching behaviour. In (d) the differential conductance d$I$/d$V$ is shown in the top insert. The temperature is 1.5 K.%
}%
  \label{anisochange}%
\end{figure}

The most interesting characteristic of the device is shown in figure~\ref{anisochange}, where the magnetoresistance measurements of the constriction for two different thermal cycles are shown in (a) and (c), along with their respective zero field current-voltage ($I$-$V$) characteristics, (b) and (d). The exponential form of the $I$-$V$ indicates tunnelling type conductivity. We partially account for the differences in resistance at $B = 0$ T of the two cases being due to local fluctuations of electrostatic potential during cool down, resulting in different preferred conduction paths and also the thermal cycling potentially causing physical changes to the very sensitive nano-contact region \cite{shi_temperature_2007}. In figure~\ref{anisochange}(a) we see that with field ${\bf B} \parallel {\bf z}$ is the high resistance state, which is the usual behaviour for these materials \cite{giddings_large_2005}.
In (c) this is now reversed, and  ${\bf B} \parallel {\bf z}$ is the low resistance state.
It is also worth pointing out that the hysteretic  ${\bf B} \parallel {\bf z}$ magnetoresistance in (c) is over 1300\%, which is comparable to the \ac{MR} effects seen in vertical \ac{TAMR} devices \cite{ruster_very_2005}. However, although the ${\bf B} \parallel {\bf z}$ magnetoresistance has changed dramatically, that for the other orientations shows much smaller changes and ${\bf B} \parallel {\bf y}$ remains a higher resistance state than ${\bf B} \parallel {\bf x}$.

\begin{figure}
  \centering
  \includegraphics*[width=0.45\textwidth]{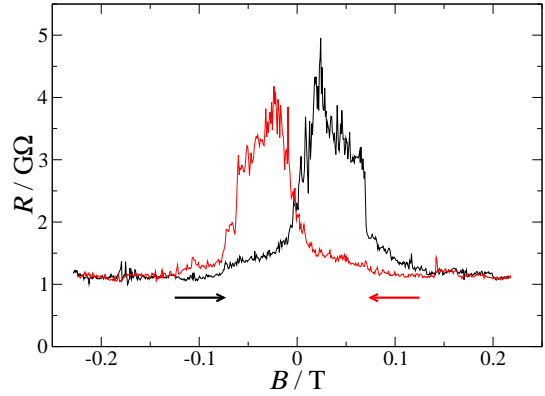}
  \caption{Magnetotransport measurement with ${\bf B} \parallel {\bf z}$. The four-point excitation is 40 mV and the temperature is 1.5 K. The black curve is for increasing field and the red curve is for decreasing field.}
  \label{notTMR}
\end{figure}

We will now briefly consider the \acl{TMR} effects previously reported in other (Ga,Mn)As nanoconstriction devices \cite{ruster_very_2003,schlapps_transport_2006}. They are of particular interest as those devices contain nanoconstrictions comparable in size to the one reported in this paper. In those devices there is a (Ga,Mn)As island, several orders of magnitude larger than the \ac{CBAMR} nano-islands, separated from (Ga,Mn)As leads by a pair of tunnel barriers, that, is a pair of nanoconstrictions. Large spin-valve effects were seen, and the explanation for this was that the island and the leads would have different coercive fields due to shape anisotropy, and so this could result in either parallel or antiparallel alignment of magnetisation between the island and the leads as the field was swept. Parallel alignments were associated with a low resistance state, while an antiparallel alignment was associated with the high resistance state \cite{ruster_very_2003}.

In figure~\ref{notTMR} the magnetoresistance trace is shown for a case with the single nanoconstriction when the field is swept with ${\bf B} \parallel {\bf z}$. Without reference to the other field orientations, the signal appears to be of a qualitatively similar nature to the spin-valve effect that a \ac{TMR} device would exhibit. However, in this case the device only contains a single constriction. The leads either side are otherwise identical and as such should have identical coercive fields. Therefore, the magnetisation of the (Ga,Mn)As either side of the constriction should remain parallel. This, therefore, precludes a \ac{TMR} mechanism, and suggests that the magnetoresistance effect is a property of the transport across or within the nanoscale area of constriction itself. Note that the mean free path of (Ga,Mn)As is less than 1 nm at temperatures lower than 1 K \cite{edmonds_hall_2002,sorensen_effect_2003} and the 5 nm film behaves as a 3D system. This highlights the difficulty in analysing \ac{TMR} transport data in (Ga,Mn)As devices containing nanoconstrictions, as there is otherwise nothing to distinguish true spin-valve behaviour from that seen in figure~\ref{notTMR}.

\begin{figure}
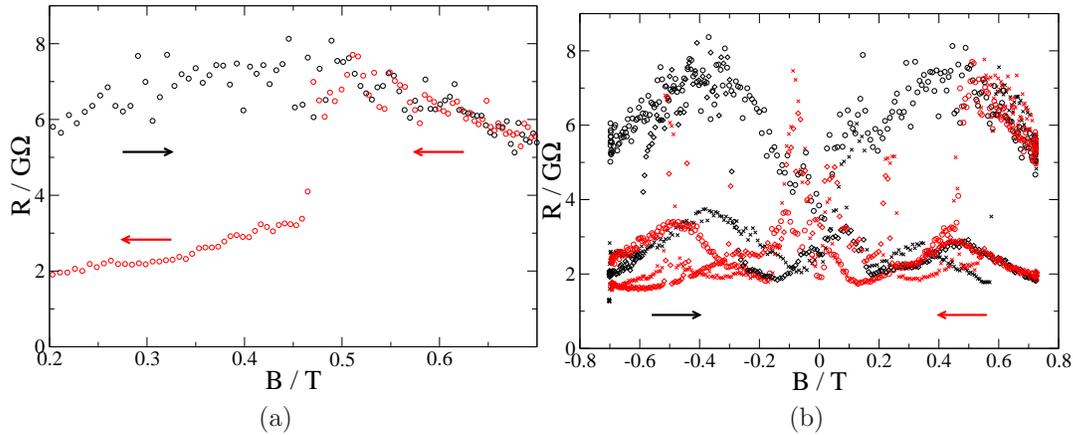

  \centering%
  \subfloat[]%
    {\includegraphics[width=0.45\textwidth]{FIGURES/fig_multistable_focus}}%
  \subfloat[]%
    {\includegraphics[width=0.45\textwidth]{FIGURES/fig_multistable2}}%
  \caption{Magnetotransport measurements across the constriction with the magnetic field
in-plane at $45\degree$ to the direction of current, showing switching events. The excitation across the device is 40 mV and the temperature is 1.5 K. (a) focuses on a single switching event and (b) shows three sequential sweeps superimposed, highlighting the various resistance states. Black points represent increasing field and red points decreasing field.}%
  \label{bistable}
\end{figure}

Further insight into the unusual behaviour of this device is shown in figure~\ref{bistable}(a). This shows measurements with the field at $45\degree$ to the current, that is, along one of the [110] axes. Switching behaviour was observed during the measurement, whereby the sample changed between high and low resistance states, with the switching occurring on a time scale from several minutes. By overlaying several consecutive field sweeps, as shown in figure~\ref{bistable}(b), it appears that the switching is occurring between a high resistance state and several similar low resistance states. This behaviour is very reminiscent of early work in \ac{SET} structures, where background charge noise would strongly feature in measurements, with telegraphic switching between two or more states \cite{zorin_background_1996}.

In a traditional 
\ac{SET} structure the resistance oscillates as a function of the gate potential on the island, leading to the so-called Coulomb diamonds. In the structure studied in this paper there is no gate and the potential of the nanoconstriction and any nano-islands will be at an arbitrary level, depending on local electrostatic conditions which vary over different thermal cycles.
We see that the charge trapping causes large changes in the resistance in the form of multistable telegraphic switching, strongly suggesting that the movement of localised charge around the tunnelling region is changing the local potential. This is similar to the effect of a gate in changing the local potential \cite{wunderlich_coulomb_2006}, and demonstrates the great sensitivity the device has to such changes.

A key feature of \ac{SET} devices is the Coulomb staircase current-voltage characteristic \cite{matsumoto_room_1996,smith_gate_1997}, whereby discrete steps in the conductivity occur as the applied voltage is increased. This effect is due to the increasing bias overcoming the charging energy of the island, which increases by one the quantised number of charges on the island. Both the points of inflection in the $I$-$V$ characteristic and the peaks in the differential conductance shown in Figure~\ref{anisochange}(d) are reminiscent of this. Disorder and multiple islands could be used explain the blurring of the steps, if this really were a Coulomb staircase. The theoretical simulation of \ac{TAMR} in this type of lateral geometry \cite{giddings_large_2005} only predicts \ac{MR} of up to about 50\%. In agreement with the conclusion of \cite{ciorga_tamr_2007}, \ac{TAMR} alone can not account for the huge \ac{MR} effects seen in this device. Taking these factors together, a Coulomb blockade based transport mechanism seems to provide a better explanation of the observed behaviour.

We therefore conclude that the dominant contribution to the \ac{MR} arises from the \ac{CBAMR} mechanism \cite{wunderlich_coulomb_2006,fernandez-rossier_anisotropic_2006}. In an effect analogous to the application of an electric field to the Coulomb blockade nano-island, during different thermal cycles the electrostatic configuration of the constriction can change dramatically resulting in large changes in the \ac{AMR} observed. We have seen that there is a strong link between the form of the in-plane \ac{AMR} of the bulk material and that of the constriction, and we account for this though the dominance of the magnetocrystalline component of the \ac{AMR}. When the field is rotated out-of-plane the shift in the chemical potential is expected to be much larger due to the strong out-of-plane anisotropies inherent in the thin films. This results in the extremely large \ac{MR} effects observed in the constriction with the field in an out-of-plane configuration. Also, one then expects the greatest sensitivity to charge fluctuation for the ${\bf B} \parallel {\bf z}$ direction, as is observed.

\section{Conclusion}

To conclude, we have demonstrated anisotropic switching behaviour in a (Ga,Mn)As thin film nanoscale device, similar to those that have been of recent interest, which results in magnetoresistance effects up to $\sim 1300$\%. We have considered the diverse magnetoresistance effects that can be exhibited in these types of structures, and discussed this result in this context. By framing the phenomenology in terms of \ac{CBAMR} we provide a likely explanation for these effects, and show how the bulk anisotropies of the material control transport behaviour in the tunnelling regimes of nanostructured devices.

\ack We thank C Marrows for the helpful discussion. We acknowledge support from EU Grant IST-015728, from UK Grant GR/S81407/01, from CR Grants 202/05/0575, 202/04/1519, FON/06/E002, AV0Z1010052, and LC510.

\acrodef{MR}{magnetoresistance}
\acrodef{AMR}{anisotropic magnetoresistance}
\acrodef{NAMR}{normal \ac{AMR}}
\acrodef{CBAMR}{Coulomb blockade anisotropic magnetoresistance}
\acrodef{DMS}{dilute magnetic semiconductor}
\acrodef{PMMA}{poly(methyl methacrylate)}
\acrodef{TAMR}{tunnelling \ac{AMR}}
\acrodef{TMR}{tunnelling \ac{MR}}
\acrodef{SET}{single electron transistor}
\acrodef{SOC}{spin-orbit coupling}
\acrodef{QD}{quantum dot}

\section*{References}
\bibliographystyle{iopart-num}
\bibliography{ref_singleconst}
\end{document}